# One-Dimensional Organometallic V-Anthracene Wire and Its B-N Analogue: Efficient Half-Metallic Spin Filters

Prakash Parida,[a] Anasuya Kundu [b] and Swapan K. Pati [ac]

**Using density functional theory, we have investigated the structural, electronic and magnetic properties of infinitely**
10 **periodic organometallic vanadium-anthracene ( [V$_2$Ant]$_\infty$) and [V$_4$(BNAnt)$_2$]$_\infty$ ( where BNAnt is B-N analogue of anthracene ) for their possible application in spintronics. From our calculations, we find that one-dimensional [V$_2$Ant]$_\infty$ and [V$_4$(BNAnt)$_2$]$_\infty$ wires exhibit robust ferromagnetic half-metallic**
15 **and metallic behavior, respectively. The finite sized V$_6$Ant$_2$ and V$_6$(BNAnt)$_2$ clusters are also found to exhibit efficient spin filter properties when coupled to graphene electrodes on either side.**

Understanding of charge transport through low dimensional electronic system has got tremendous impetus in recent years
20 because of their huge potential in advanced electronic devices [1]. A new discipline which exploits both the electronic and spin degrees of freedom, called spintronics [2], has come forth as a promising field at both fundamental and applied levels [3]. Half-metals, a new class of compounds which exhibit large carrier
25 spin polarizations, show promising features for spintronics applications, owing to the coexistence of the metallic nature of conduction in one spin orientation and an insulating nature for the other spin orientation [4]. Half metallic (HM) behavior has been reported in three-dimensional materials, such as
30 $CrO_2$, manganite perovskites and two-dimensional materials such as graphene nano ribbons [5].

However, with conventional microelectronics moving towards nano-electronics, theoreticians and experimentalists
35 are enthusiastically pursuing the search for one-dimensional organometallic materials which have huge potential in advanced spintronic devices. To this regard, after the discovery of ferrocene, transition metal atom (TM)-cyclopentadienyl, TM-benzene, TM-borazine, TM-pentalene
40 multideckers and 1-D TM-benzimidazole have been extensively studied both theoretically and experimentally by many groups [6–8]. It would also be interesting to investigate if aromatic compounds larger than benzene can make similar multideckers, since they can hold more than one transition
45 metal atom in the sandwiched region. In additon, V-naphthalene and V-anthracene clusters have been synthesized by reacting vanadium vapour with naphthalene and anthracene in gas phase [9]. Although vanadium-naphthalene systems have been extensively studied theoretically, there is little
50 information about the vanadium-anthracene system [10]. For a detailed understanding, we consider here vanadium-anthracene nanowire and compare its electronic, magnetic and device characteristics with their B-N analogues, [V$_4$(BNAnt)$_2$]$_\infty$ (where BNAnt corresponds to anthracene with every C-C
55 bonds replaced by B-N). Note that, BNAnt is isoelectronic with Ant and we attempt here for the first time to rationalize their properties when physisorbed with V to form multidecker complexes.

60 All our calculations have been performed with density functional theory (DFT)-based spin-polarized first-principles approach as implemented in the SIESTA package [11]. A double zeta polarized (DZP) basis set with the polarization of orbitals has been included for all the atoms. The semi-core 3p orbital
65 has been included in valence orbital for transition metal atoms. The Perdew, Burke and Ernzerhof (PBE) version of the generalized gradient approximation (GGA) functional is adopted for exchange-correlation [12]. A real space mesh cut-off of 400 Ry is used for calculating the electronic structure of all
70 systems. We have also carried out calculations on spin states, magnetic moments and Density of states (DOS) for both the systems using B3LYP [13] functional as implimented in CRYSTAL06 code [14]. Previous studies on one-dimensional systems have shown that the GGA results compare fairly well
75 with hybrid functionals such as B3LYP [7, 15]. In our case too, we find qualitative agreement between the two functionals. For one dimensional sandwich molecular wires (SMWs), a unit cell of 20 × 20 × c Å$^3$ ( 20 × 20 × 2c Å$^3$ ), where c is the distance between two neighbouring Ant (BNAnt) units, has
80 been used. To understand the magnetic interactions among the transition metal atoms in SMWs, we consider a supercell which contains four vanadium atoms and two units of Ant (or BNAnt). We have used a k-mesh of 1 × 1 × 40 for periodic calculation. To understand the transport properties, we have
85 calculated the transmission function, T(E), at zero bias using the non-equilibrium Green function methodology extended for spin-polarized systems [16].

We first present the results of the structural analysis for
90 [V$_2$Ant]$_\infty$ and [V$_4$(BNAnt)$_2$]$_\infty$ SMWs. The optimized geometries are presented in Fig. 1. It is to be noted that [V$_4$(BNAnt)$_2$]$_\infty$ can adopt two eclipsed structures, where the difference between the two structures is reflected due to the relative position of the B atoms with respect to the N atoms in
95 the consecutive BNAnt unit. Interestingly, the eclipsed conformer with the B atoms on top of the N atoms in the consecutive BNAnt unit (Eclipsed-1) is stabilized over the other eclipsed conformer where the B atoms are on top of

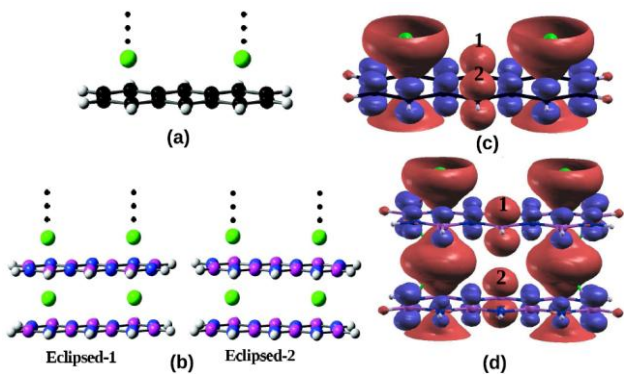
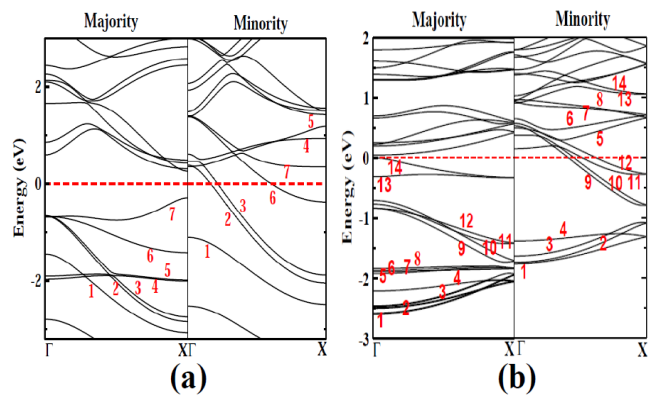

Fig. 1: Optimized Geometry of (a) $[V_2Ant]_\infty$. And (b) $[V_4(BNAnt)_2]_\infty$. Spin density map for (c) $[V_2Ant]_\infty$. and (d) $[V_4(BNAnt)_2]_\infty$. Code: red isosurface shows for the positive magnetic moment, while blue shows for the negative magnetic moment.

Fig. 2. Spin resolved band structure of (a) $[V_2Ant]_\infty$ and (b) $[V_2BNAnt]_\infty$. The plot is scaled for $E_F$ to lie at 0 eV.

each other (Eclipsed-2). We note that, in the former, the B atoms are on top of the N atoms, thereby favoring charge transfer stabilization. Thus, we consider the Eclipsed-1 geometry for further studies. The interplanar separation between the BNAnt units (in Eclipsed-1 geometry) is found to be 3.66 Å in $[V_4(BNAnt)_2]_\infty$ while this separation between anthracene unit is 3.56 Å in $[V_2Ant]_\infty$. To estimate the stability of $[V_2Ant]_\infty$ and $[V_4(BNAnt)_2]_\infty$ SMWs, we calculate the binding energy per V atom, defined as

$BE([V_2Ant]_\infty) = ( E(Ant) + 2E(V) – E([V_2Ant]_\infty))/2$

$BE([V_4(BNAnt)_2]_\infty)=(2E(BNAnt)+4E(V)-E([V_4(BNAnt)_2]_\infty)/4$

where $BE([V_2Ant]_\infty)$ ( or $BE([V_4(BNAnt)_2]_\infty)$ ) is the binding energy for $[V_2Ant]_\infty$ ( or $[V_4(BNAnt)_2]_\infty$ ), E(V) is the energy of the isolated V atom, and E(Ant) (or E(BNAnt) ) is the energy of isolated Ant ( or BNAnt ) unit. The binding energy for $[V_2Ant]_\infty$ and $[V_4(BNAnt)_2]_\infty$ are found to be 7.0 eV and 4.59 eV respectively. It is very clear that, binding strength of V-Ant system is stronger than that of V-BNAnt system. This is due to the fact that, in the anthracene molecule, the electrons are fully delocalized and thus the whole electron cloud can interact strongly with metal ligand, while in BNAnt, the charges are localized on B and N atoms in a charge transfer state, thereby interacting weakly with metal ligand.

To probe the magnetic interaction, we consider a supercell containing four V atoms and two anthracene (or BNAnt) units. We calculate the energy of each system with four different types of alignments of local magnetic moment on four metal atoms, i.e, case(a) FM alignment, case(b) AFM alignment, case(c) interlayer FM with intralayer AFM and case(d) interlayer AFM with intralayer. We find that the two systems are ferromagnetically (case a) stabilized for both the functionals, GGA and B3LYP (see the Table 1). Although the energy difference between different spin configurations are not quantitatively same for different functionals, the fundamental results remain same for both the functionals. To understand the magnetic interactions, we calculate the net magnetic moment of the unit cell per V atom and this value is found to be 2 $\mu_B$ and 1.98 $\mu_B$ for $[V_2Ant]_\infty$ and $[V_4(BNAnt)_2]_\infty$, respectively and almost remain same for the B3LYP functional as well. From the spin density map (see Fig1), it can be seen that V atoms carry large magnetic moments for both the systems. In fact, we find a positive magnetic moment of +2.18 $\mu_B$ on each V atom while each anthracene molecule has a small negative magnetic moment (-0.18$\mu_B$), which is distributed over the entire structure. For the case of $[V_4(BNAnt)_2]_\infty$, this value is +2.16 $\mu_B$ on each V atom and -0.18 $\mu_B$ on BNAnt molecule. The important point to note here is that, the magnetic moment is not distributed uniformly over the C atoms in $[V_2Ant]_\infty$. We also find that, the magnetic moment is not uniform over B atoms and N atoms in $[V_4(BNAnt)_2]_\infty$. Interestingly, a small but negative magnetic moment is found over all C atoms in $[V_2Ant]_\infty$ and over all B and N atoms in $[V_4(BNAnt)_2]_\infty$ except the middle C, B and N atoms (see the marks 1 and 2 in Fig. 1(c) and Fig. 1(d), over which a finite positive magnetic moment is present. It is well known that the middle C atoms at the marked positions 1 and 2 in anthracene molecule is highly reactive unlike the atoms at other positions. It is precisely due to this, the atoms at these special positions posses a finite positive moment on them. We also note that N atoms contribute relatively more to the negative moment (-0.034 $\mu_B$ per N) when compared to the B atoms (-0.025 $\mu_B$ per B) except at the marked positions 1 and 2, where B atoms (+0.23 $\mu_B$) contribute much more to the positive moment when compared to N atoms (+0.014 $\mu_B$ ). It is also clear that the moment per unit cell per V atom in both the systems $[V_2Ant]_\infty$ and $[V_4(BNAnt)_2]_\infty$ are approximately same.

To understand the origin of finite magnetic moment, we present the spin-polarized band structure for $[V_2Ant]_\infty$ and $[V_4(BNAnt)_2]_\infty$ in Fig. 2. Note that, there are two(four) d-derived bands for each of five d orbitals since there are two(four) vanadium atoms in the primitive cell of $[V_2Ant]_\infty$ ($[V_4(BNAnt)_2]_\infty$). Furthermore, we note that, although isolated V atom has the valence electronic configuration $3d^34s^2$, due to strong hybridization, the vanadium 4s levels are shifted above $E_F$ making the effective valence configuration $3d^54s^0$. Also, the states derived from $d_{xz}$ and $d_{yz}$ states are pushed above the Fermi level and remains unoccupied for both types of spin, making no appreciable contribution to the magnetic moment. Since the unit cell of $[V_2Ant]_\infty$ contains two V atoms, i.e, ten d electrons, seven majority spin electrons completely fill the seven bands, leading to the opening up of a semiconducting band gap of 0.54 eV at X point in the majority spin channel. On the other hand, out of the remaining three electrons, one electron fill the minority spin band (band 1 in Fig. 2(a)) and other two minority electrons partially fill three bands (bands 2, 3 and 6 in Fig. 2(a)). In fact, due to such partial filling, both the bands cross the $E_F$, closing the minority band gap. Thus, such a filling results in a net magnetic moment of 4 $\mu_B$ per unit cell of $[V_2Ant]_\infty$. The most important point here is that, we have a coexistence of the metallic and semiconducting nature for electrons in the minority and majority spin channels, respectively, leading to a half-metallic (HM) behavior for the $[V_2Ant]_\infty$ with a net magnetic moment of 4 $\mu_B$. Interestingly, the flat portions of the bands 4 and 5 are mostly $d_{z^2}$ derived states (a localized state) while the dispersive portions are $d_{x^2-y^2}$ derived states and $d_{xy}$ derived states. The band 7 is mainly $d_{x^2-y^2}$ derived states with very little contribution from $d_{z^2}$ orbitals. The numbering of bands has been done by considering the symmetry of orbitals involved in the bands. For the case of $[V_4(BNAnt)_2]_\infty$, the net moment per unit cell is 8 $\mu_B$. Since the unit cell contains four V atoms, 14 majority spin electrons fill 14 bands. Interestingly, the top of the valence band (band 14) just crosses the Fermi level and the bottom of conduction band remains closer to the Fermi level for the majority spin. On the other hand, out of remaining six d electrons, four minority electrons completely fill four minority bands and the other two electrons partially fill four bands crossing the Fermi level. Thus $[V_4(BNAnt)_2]_\infty$ wire is metallic in nature. As can be seen from the Fig. 2(b), the bands 5, 6, 7, 8, 13, and 14 are completely filled for the majority spin channel while completely empty for the minority spin channel, making contribution of 6 $\mu_B$ to the net magnetic moment per primitive cell. The remaining moment of 2 $\mu_B$ is contributed from four partially filled minority bands (bands 9, 10, 11, and 12 in Fig. 2(b)). In fact we would like to stress that, due to the large moment, the $[V_2Ant]_\infty$ and $[V_4(BNAnt)_2]_\infty$ wires would be much more useful for molecular magnets compared to their benzene and borazine analogues.

Table 1: Energy (in meV) for different spin configurations for $[V_2Ant]_\infty$, and $[V_4(BNAnt)_2]_\infty$ systems with different functionals. Case (a) corresponds to FM; case (b) corresponds to AFM; case (c) corresponds to interlayer FM with intra layer AFM; and case (d) corresponds to interlayer AFM with intra layer FM. All the energies are scaled with respect to the energy of the FM case.

| System and Method | Case (a) | Case (b) | Case (c) | Case (d) |
|---|---|---|---|---|
| $[V_2Ant]_\infty$ | | | | |
| GGA | 0.000 | 910 | 240 | 660 |
| B3LYP | 0.000 | 707 | 160 | 382 |
| $[V_2BNAnt]_\infty$ | | | | |
| GGA | 0.000 | 600 | 520 | 320 |
| B3LYP | 0.000 | 440 | 280 | 136 |

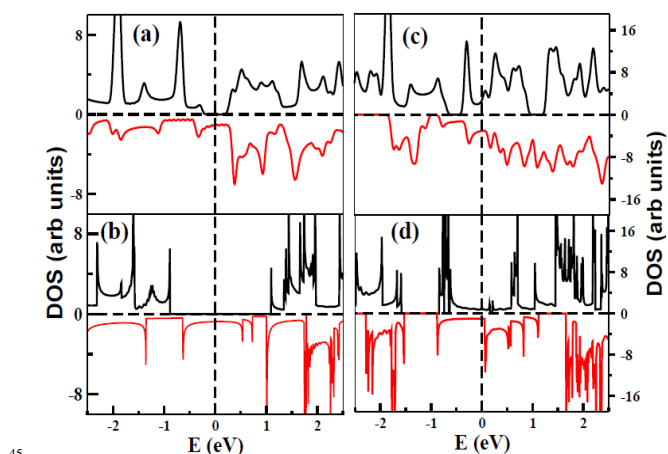

Fig. 3: Spin resolved DOS for $[V_2Ant]_\infty$ with (a) GGA and (b) B3LYP. The same for $[V_4(BNAnt)_2]_\infty$ with (c) GGA and (d) B3LYP. The plot is scaled for $E_F$ to lie at 0 eV. Code: solid black and red lines show the DOS for majority and minority spin channels, respectively.

For further understanding of the energy level structure leading to half-metallic and metallic behaviour for our systems, we present the DOS plots in Fig. 3. For both GGA and B3LYP, as can be seen, at Fermi level ($E_F$), a gap opens for majority spin channel and the finite density of states for the minority spin channel leading to the half metallicity for $[V_2Ant]_\infty$, while $[V_4(BNAnt)_2]_\infty$ shows metallic behavior with finite density of states for both the spin channels. Both the methods show qualitatively similar behavior with same conclusion about the overall electronic states of the systems.

We now investigate the transport properties of the finite size metal composites, which can be experimentally synthesized, and probe them for possible device applications. The isostructural similarities between graphene and anthracene (or BNAnt) along with the strong adsorption of V atoms on graphene facilitates the stabilization and deposition of these $V_{2n}Ant_{n+1}$ and $V_{2n}(BNAnt)_{n+1}$ clusters on graphene electrodes [17]. Here, we give details of our calculation on transport properties of $V_6Ant_2$ and $V_6(BNAnt)_2$ finite sized clusters adsorbed on graphene sheet electrodes on either side. The Electrode-Molecule-Electrode (EME) system is modeled as Graphene-$V_2(V_2Ant_2)V_2$-Graphene and Graphene-$V_2(V_2BNAnt_2)V_2$-Graphene. Note that, we have terminated the cluster by V atoms to improve the binding to the graphene electrodes. We have used edge-passivated graphene to model the electrodes, with each electrode consisting of 90 C atoms and 26 H atoms. Our spin polarized calculations suggest that the central and the outermost V atoms posses the magnetic moment of 2.52 (2.35) $\mu_B$ and 2.4 (2.55) $\mu_B$ per V atom for Graphene-$V_2(V_2Ant_2)V_2$-Graphene (Graphene-$V_2(V_2BNAnt_2)V_2$-Graphene), respectively and the local magnetic moments are coupled ferromagnetically with each other.

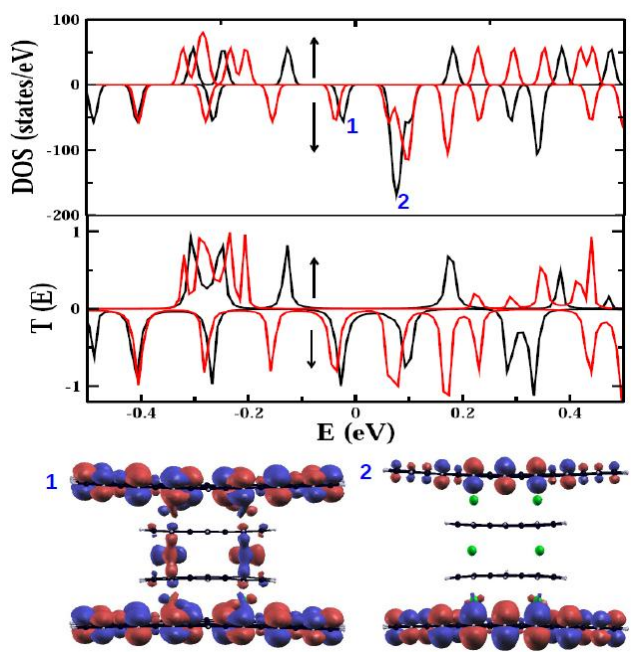

**Fig. 4:** Spin resolved DOS and T(E) plot for Graphene-$V_2(V_2Ant_2)V_2$-Graphene (black line) and Graphene-$V_2(V_2BNAnt_2)V_2$-Graphene (red line). The plot is scaled for $E_F$ to lie at 0 eV. Important orbitals have been labeled and the corresponding orbitals have been plotted. HOMO and LUMO of Graphene-$V_2(V_2Ant_2)V_2$-Graphene are labeled as 1 and 2, respectively.

In Fig. 4, we plot the spin polarized DOS and transmission, T(E), for the two systems. From the DOS plot for Graphene-$V_2(V_2Ant_2)V_2$-Graphene (Graphene-$V_2(V_2BNAnt_2)V_2$-Graphene), we find that the highest occupied molecular orbital (HOMO) is 0.12 (0.20) eV below $E_F$ in the majority spin channel, while it is 0.02 (0.03) eV below the $E_F$ in the minority spin channel. On the other hand, the lowest unoccupied molecular orbital (LUMO) is 0.18 (0.22) eV above the $E_F$ in the majority spin channel, while it is 0.07 (0.06) eV above the $E_F$ for the minority spin channel. Since the low-energy molecular orbitals around the Fermi level govern the transport phenomenon, we restrict our discussion of T(E) to the HOMO and LUMO of the majority and minority spin channels, respectively. As can be seen from T(E) plot, in the low bias window around the Fermi level, only the minority spin electrons take part in the spin transport while majority spin electrons do not contribute to the spin transport for both the systems. It is also clear that within a very small energy window of ±0.05 eV around the Fermi energy, only HOMO contributes to the transport for minority spin channel in both the systems. In fact, in quantitative terms, Graphene-$V_2(V_2Ant_2)V_2$-Graphene shows strong transmission peak of 1.0 compared to that of 0.8 for Graphene-$V_2(V_2BNAnt_2)V_2$-Graphene. Interestingly, although the minority spin electrons are responsible for the transport in the low bias window around Fermi level, in particular, the LUMO of minority spin channel in Graphene-$V_2(V_2Ant_2)V_2$-Graphene does not show any strong transmission peak, while the corresponding HOMO shows a stong transmission peak. This can be clearly understood from the analysis of the orbital plots of the HOMO and LUMO of the minority spin channels (see the Fig. 4 ) in Graphene-$V_2(V_2Ant_2)V_2$-Graphene. Note that, while LUMO of the minority spin channel is delocalized over whole EME system, the corresponding LUMO is quite localized over the entire EME system, thereby hindering the transport.

To conclude, we have critically examined the electronic and magnetic properties of the $[V_2Ant]_\infty$ and $[V_4(BNAnt)_2]_\infty$. We predict that $[V_2Ant]_\infty$ and $[V_4(BNAnt)_2]_\infty$ 1-D wires are ferromagnetic half-metal and metal, respectively. We also find that finite sized $V_6Ant_2$ and $V_6(BNAnt)_2$ clusters exhibit efficient spin filter properties when coupled to graphene electrodes on either side. Owing to better structural stability and robust electronic behaviour, we conjecture that the $V_{2n}Ant_{n+1}$ and $V_{2n}(BNAnt)_{n+1}$ systems would act as efficient spin filters for advanced spintronics applications.

PP acknowledges the CSIR for a research fellowship and SKP acknowledges research support from CSIR and DST, Government of India and AOARD, US. Airforce, for the research grants [a] *Theoretical Sciences Unit, Jawaharlal Nehru Center for Advanced Scientific Research, Jakkur Campus, Bangalore-560064, India. Email: pati@jncasr.ac.in*
[b] *Theoretical Condensed Matter Physics Division, Saha Institute of Nuclear Physics, Bidhannagar, Kolkota-700064, India*
[c] *New Chemistry Unit, Jawaharlal Nehru Centre For Advanced Scientific Research, Jakkur Campus, Bangalore-560064, India.*

1. A. Nitzan and M. A. Ratner, *Science*, 2003, **300**, 1384; W. B. Davis, W. A. Svec, M. A. Ratner and M. R. Wasielewski, *Nature*, 1998, **396**, 60; R. G. Endres, D. L. Cox and R. R. P. Singh, *Rev. Mod. Phys.*, 2004, **76**, 195.
2. T. Rueckes, K. Kim, E. Joselevich, A. R. Rocha, V. M. Garcia-suarez, S. W. Bailey, C. J. Lambert, J. Ferrer, and S. Sanvito, *Nature Materials*, 2005, **4**, 335; G. A. Prinz, *Science*, 1998, **282**, 1660.
3. I. Zutic, J. Fabian and S. D. Sarma, *Rev. Mod. Phys.*, 2004, **76**, 323; S. A. Wolf, D. D Awschalom, R. A. Buhrman, J. M. Daughton, S. Von Molnar, M. L. Roukes, A. Y. Chtchelkanova, and D. M. Treger, *Science*, 2004, **294**, 1488.
4. R. A. de Groot, F. M. Mueller, P. G. van Engen and K. H. J. Buschow, *Phys. Rev. Lett.*, 1983, **50**, 2024; M. I. Katsnelson, V. Yu. Irkhin, L. Chioncel, A. I. Lichtenstein, R. A. de Groot, *Rev. Mod. Phys.*, 2008, **80**, 315.
5. Y. -W. Son, M. L. Cohen and S. G. Louie, *Nature*, 2006, **444**, 347; S. Dutta, A. K. Manna and S. K. Pati, *Phys. Rev. Lett.*, 2009, **102**, 096601; S. Dutta and S. K. Pati, *J. Phys. Chem. B*, 2008, **112**, 1333.
6. D. B. Jacobson, B. S. Freiser, *J. Am. Chem. Soc.*, 1984, **106**, 3900; L. Zhou, S. -W. Yang, M. -F. Ng, M. B. Sullivan, V. B. C. Tan and L. Shen, *J. Am. Chem. Soc.*, 2008, **130**, 4023; L. Shen, S. -W. Yang, M. -F. Ng, V. Ligatchev, L. Zhou, and Y. Feng, *J. Am. Chem. Soc.*, 2008, **130**, 13956; R. Liu, S. -H. Ke, H. U. Baranger, and W. Yang, *Nano Lett.*, 2005, **5**, 1959; R. Liu, S. -H. Ke, W. Yang, H. U. Baranger, *J. Chem. Phys.*, 2007, **127**, 141104; S. S. Mallojosyula, P. Parida and S. K. Pati, *J. Mat. Chem.*, 2009, **19**, 1761; L. Zhu and J. Wang, *J. Phys. Chem. C*, 2009, **113**, 8767.
7. V. V. Maslyuk, A. Bagrets, V. Meded, A. Arnold, F. Evers, M. Brandbyge, T. Bredow and I. Mertig, *Phys. Rev. Lett.*, 2006, **97**, 097201.

8   R. Pandey, B. K. Rao, P. Jena, and M. A. Blanco, *J. Am. Chem. Soc.*, 2001, **123**, 3799; A. K. Kandalam, B. K. Rao, P. Jena, R. Pandey, *J. Chem. Phys*., 2004, **120**, 10414; T. Yasuike, A. Nakajima, S. Yabushita, and K. Kaya, *J. Phys. Chem. A*., 1997, **101**, 5360; K. Miyajima, A. Nakajima, S. Yabushita, M. B. Knickelbein and K. Kaya, *J. Am. Chem. Soc*., 2004, **126**, 13202; K. Miyajima, S. Yabushita, M. B. Knickelbein and A. Nakajima, *J. Am. Chem. Soc*., 2007, **129**, 8473; X. Wu and X. C. Zeng, *J. Am. Chem. Soc*., 2009, **131**, 14246; S. S. Mallajosyula and S. K. Pati, *J. Phys. Chem. B*, 2008, **112**, 16982; S. S. Mallajosyula and S. K. Pati, *J. Phys. Chem. B*, 2007, **111**, 13877; X. Zhang, J. Wang, Y. Gao, X. C. Zeng, *ACS Nano*, 2008, **3**, 537.
9   T Kurikawa et al., *Organometallics*, 1999, **18**, 1430.
10  H. S. Kang, *J. Phys. Chem. A*, 2005, **109**, 9292; L. Wang et. al., *Nano Lett*., 2008, **8**, 3640.
11  J. M. Soler, *J. Phys.: Condens. Matter*, 2002, **14**, 2745 .
12  J. P. Perdew, K. Burke, M. Ernzerhof, *Phys. Rev Lett*., 1996, **77**, 3865.
13  C. Lee, W. Yang, R. G. Parr, *Phys. Rev. B*, 1988, **37**, 785; A. D. Becke, *J. Chem. Phys*., 1993, **98**, 5648.
14  R. Dovesi, V. R. Saunders, C. Roetti, R. Orlando, C. M. Zicovich-Wilson, F. Pascale, B. Civalleri, K. Doll, N. M. Harrison, I. J. Bush, Ph. D'Arco and M. Llunell, 2006, *Crystal 2006 User's Manual* (Torino: University of Torino).
15  H. Xiang, J. Yang, J. G. Hou and Q. Zhu, *J. Am. Chem. Soc.,* 2006, **128**, 2310
16  S. Datta, Electronic Transport in Mesoscopic Systems, *Cambridge University Press: Cambridge*, 1995; S. Lakshmi, S. Dutta and S. K. Pati, *J. Phys. Chem. C*, 2008, **112**, 14718; S. Sengupta, S Lakshmi and S. K. Pati, *J. Phys.: Condens. Matter,* 2006, **18**, 9189.
17  M. Koleini, M. Paulsson and M. Brandbyge, *Phys. Rev Lett*., 2007, **98**, 197202 ; C. -K. Yang, J. Zhao and J. P. Lu *Nano Lett*., 2004, **4,** 561.